\documentclass[11pt,twocolumn]{article}
\usepackage{epsfig,epstopdf}
\usepackage[utf8x]{inputenc}
\usepackage{amsmath,graphicx}
\usepackage{color}

\newcommand{\bn}{{\bar n}}

\newcommand{\nslash}{{\not \!n}}
\begin{document}
\title{The $T$-Wilson Line}

\author{Ahmad Idilbi and Ignazio Scimemi\\
Departamento de F\'isica Te\'orica II,
Universidad Complutense de Madrid (UCM),\\
28040 Madrid, Spain}
\maketitle
\begin{abstract}
We argue that soft-collinear effective theory, as is currently formulated, lacks an ingredient which is essential for obtaining a gauge invariant definitions of certain non-perturbative matrix elements relevant for semi-inclusive high-energy processes. Those processes and the companion cross-sections are to be analyzed and measured at the Large Hadron Collider. The missing feature is a gauge link constructed only from the transverse component of the gluon field denoted by the ``$T$-Wilson line''. It is neither soft nor collinear. As such it is a new feature of the theory.
\end{abstract}

\section{Introduction}
Soft-Collinear effective theory (SCET) \cite{SCET1,SCETf,Bauer:2002nz} has been established in recent years as the
most suitable and economic framework to obtain factorization theorems for high-energy processes. This is especially
important when the Large Hadron Collider (LHC) is about to be operated at its highest possible energies.

The success of SCET in deriving such factorization theorems relies on different factors, however here and for brevity,
we concentrate on two of them. First is the fact that the soft and collinear modes (which capture the infra-red singularities
in perturbartive QCD)
are decoupled at the level of SCET Lagrangian and operators. The second is that the fundamental building blocks
of SCET enable us to derive, from first principles (of SCET itself), a gauge invariant definitions of the various
non-perturbative matrix elements that enter into those factorization theorems.

In this letter we argue that the last statement holds only for certain class of gauges known as ``regular'' gauges
in which the gauge field vanishes at infinity in one light-cone direction such as the case in all covariant gauges.
In the case of ``singular'' gauges, like in the light-cone gauge where the the transverse components of the gluon
field do not vanish at light-cone infinity
\cite{jack} then certain modifications has to be applied for the SCET building blocks to render them gauge invariant
no matter what is the gauge fixing condition used to quantize the theory.
\section{Jet In SCET}
For concreteness let us consider the matrix element of the pure collinear
jet in SCET: $\langle 0| W_\bn^\dagger \xi_\bn| q_\bn\rangle$ which describes an incoming parton moving along
the $\bn=\frac{1}{\sqrt 2}(1,0,0)$ direction (our notation is $(+,-,\perp)$
for light-cone coordinates). The $ W_\bn^\dagger$ is the familiar SCET collinear Wilson line
and $\xi_\bn$ stands for the collinear quark field and here ``pure'' jet means that the zero-bin contribution \cite{Manohar:2006nz} (see also \cite{Idilbi:2007ff,Lee:2006nr}) has already been subtracted out. Now one can simply perform a one-loop calculation for the jet function in two different gauges, say, Feynman and light-cone gauges. In light-cone gauge we choose $A^+=0$ which renders $ W_\bn^\dagger =1$. To perform the calculation in light-cone gauge one is enforced to specify a prescription to regularize the spurious light-cone singularity stemming from the $1/k^+$ term (or the ``axial term'') in light-cone gauge. However it has been verified \cite{Bassetto:1984dq} that the Mandelstam-Leibbrandt (ML) prescription is the only valid prescription to canonically quantize QCD while preserving causality, unitarity and QCD power counting. Thus one can easily be convinced that this prescription is also the valid one to be considered in SCET.

Following the above one can demonstrate that the pure SCET
 jet function $\langle 0| W_\bn^\dagger \xi_\bn| q_\bn\rangle$ attains
different values in Feynman gauge and in light-cone gauge with the ML
prescription \cite{Idilbi:2010im}. Most noticeable is the fact that in
 the former gauge one gets, to first order in $\alpha_s$, a double pole while in the latter there is only a single one.

The last result can be understood once we recall, as mentioned above that in light-cone gauge (with $A^+=0$) and at
 infinity in light-cone direction one can still perform a gauge transformation on the quark field in the jet function.
 This transformation is built-up from the non-vanishing transverse components of the gluon field and it depends only
 on the length $r_\perp$ in the transverse space of the collinear trajectory. This gauge transformation has to be
compensated for by the existence of yet another Wilson line in the jet function. Clearly this Wilson line has to
contain only the transverse components of the gluon field and it has to be calculated at infinity in light-cone direction.
This additional Wilson line to the jet function will be denoted by $T$ and the truly gauge invariant SCET jet is given by:
\begin{align}
\label{eq:buildingblocks2}
\langle 0|T_{\bn}^\dagger W_{\bn}^\dagger\xi_\bn|q_\bn\rangle\ ,
\end{align}
where the $T$-Wilson line is
\begin{align}
 T_{\bn}^\dagger(x^+,x_\perp)={\cal P}\exp\left[i g\int_0^\infty d\tau {\bf l}_\perp\nonumber\right.\\
 \left.\cdot {\bf A}_\perp(\infty^-,x^+;{\bf l}_\perp\tau+{\bf x}_\perp)
\right]\ .
\end{align}
 The  vector $\bf l$ specifies a  path in a two-dimensional space.

 If one considers SCET in Feynman gauge (or any other covariant gauge) then the $T$-Wilson
line becomes unity since in regular gauge the gluon field vanishes at infinity in light-cone direction
thus many of the well-established SCET results are not altered by the introduction of this additional Wilson line. However
its introduction allows us now to define, gauge invariantly, quantities such as transverse-momentum parton distribution functions (TMDPDF), certain types of beam functions and any other type of non-perturbative matrix elements where the two parton fields entering the definition of such quantities are separated
 in one light-cone direction as well as in the transverse one.
\section{Application: TMDPDF In SCET}
As an example we take the TMDPDF in QCD and in SCET.
 In QCD the  gauge invariant definition  was first given in Ref.~\cite{yuan} and studied in light-cone gauge in Ref.~\cite{cs}.
In all these works it is pointed out how in QCD new kind of divergences appear
when the fields entering  the matrix elements are separated in the transverse direction.
The TMDPDF occur for instance in SIDIS  and here the complete factorization is achieved only in the presence of transverse links.
Within SCET  the TMDPDF  for a quark $q$ in a hadron $P$ with momentum $p$ can be defined with
 the use of   $\underline{\chi}$-field
\begin{align}
\underline{\chi}_\bn (y)\equiv T^\dagger_\bn (y^+,{\bf y}_\perp)W^\dagger_\bn(y) \xi_\bn (y)  , \nonumber \\
\phi_{q/P}=\langle P_\bn|\overline{\underline{\chi}}_\bn (y)\delta\left(x-\frac{n {\cal P}}{n p}\right)\delta^{(2)}(p_\perp-{\cal P}_\perp)\times\nonumber\\
\frac{\nslash}{\sqrt{2}}
\underline{\chi}_\bn (0)|P_\bn\rangle \ ,
\end{align}
where $x$ is the momentum fraction of the quark in the light-cone direction, and ${\cal P}$ is
the usual label operators in SCET.
The above definition of the TMDPDF is now gauge invariant in both classes of gauges.

We finally mention that the origin of the $T$-Wilson line in SCET (and in light-cone gauge)
 is QCD itself as is the situation with collinear Wilson line $W$ relevant in covariant gauge.
More on this and related issues will be published in a forthcoming paper \cite{future}.

\end{document}